\documentclass{llncs}

\usepackage{makeidx}  
\usepackage[T1]{fontenc}
\usepackage[latin9]{inputenc}
\usepackage{amsmath}
\usepackage{esint}
\usepackage{graphicx}
\usepackage{color}

\begin{document}

\frontmatter          

\mainmatter              

\title{Probabilistic Constraint Programming for Parameters Optimisation of Generative Models}

\titlerunning{Probability Constraint Programming}  

\author{Massimiliano Zanin, Marco Correia, Pedro A. C. Sousa and Jorge Cruz}

\authorrunning{M. Zanin et al.} 

\tocauthor{---}

\institute{NOVA Laboratory for Computer Science and Informatics, FCT/UNL, Portugal\\
\email{m.zanin@campus.fct.unl.pt, \{mvc, pas, jcrc\}@fct.unl.pt}
}

\maketitle              

\begin{abstract}
Complex networks theory has commonly been used for modelling and understanding the interactions taking place between the elements composing complex systems. More recently, the use of generative models has gained momentum, as they allow identifying which forces and mechanisms are responsible for the appearance of given structural properties. In spite of this interest, several problems remain open, one of the most important being the design of robust mechanisms for finding the optimal parameters of a generative model, given a set of real networks.
In this contribution, we address this problem by means of Probabilistic Constraint Programming. By using as an example the reconstruction of networks representing brain dynamics, we show how this approach is superior to other solutions, in that it allows a better characterisation of the parameters space, while requiring a significantly lower computational cost.

\keywords{Probabilistic Constraint Programming, Complex networks, Generative models, Brain dynamics}
\end{abstract}

\section{Introduction}
\label{sec:Intro}

The last decades have witnessed a revolution in science, thanks to the appearance of the concept of {\it complex systems}: systems that are composed of a large number of interacting elements, and whose interactions are as important as the elements themselves \cite{Anderson1972}. In order to study the structures created by such relationships, several tools have been developed, among which {\it complex networks theory} \cite{Albert2002,Newman2003}, a statistical mechanics understanding of graph theory, stands out.

Complex networks have been used to characterise a large number of different systems, from social \cite{Costa2011} to transportation ones \cite{Zanin2013}. They have also been valuable in the study of brain dynamics, as one of the greatest challenges in modern science is the characterisation of how the brain organises its activity to carry out complex computations and tasks. Constructing a complete picture of the computation performed by the brain requires specific mathematical, statistical and computational techniques. As brain activity is usually complex, with different regions coordinating and creating temporally multi-scale, spatially extended networks, complex networks theory appears as the natural framework for its characterisation.

When complex networks are applied to brain dynamics, nodes are associated to sensors ({\it e.g.} measuring the electric and magnetic activity of neurons), thus to specific brain locations, and links to some specific conditions. For instance, brain functional networks are constructed such that pairs of nodes are connected if some kind of synchronisation, or correlated activity, is detected in those nodes - the rationale being that a coordinated dynamics is the result of some kind of information sharing \cite{Bullmore2009}. Once these networks are reconstructed, graph theory allows endowing them with a great number of quantitative properties, thus vastly enriching the set of objective descriptors of brain structure and function at neuroscientists' disposal. This has especially been fruitful in the characterisation of the differences between healthy (control) subjects and patients suffering from neurologic pathologies \cite{Papo2014}. 

Once the topology (or structure) of a network has been described, a further question may be posed: can such topology be explained by a set of simple generative rules, like a higher connectivity of neighbouring regions, or the influence of nodes physical position? When a set of rules (a {\it generative model}) has been defined, it has to be optimised and validated: one ought to obtain the best set of parameters, such that the networks yielded by the model are topologically equivalent to the real ones. This usually requires maximising a function of the $p$-values representing the differences between the characteristics of the synthetic and real networks. In spite of being accepted as a standard strategy, this method presents several drawbacks. First, its high computational complexity: large sets of networks have to be created and analysed for every possible combination of parameters; and second, its unfitness for assessing the presence of multiple local minima.

In this contribution, we propose the use of {\it probabilistic constraint programming} (PCP) for characterising the space created by the parameters of a generative model, {\it i.e.} a space representing the distance between the topological characteristics of real and synthetic networks. We show how this approach allows recovering a larger quantity of information about the relationship between model parameters and network topology, with a fraction of the computational cost required by other methods. Additionally, PCP can be applied to single subjects (networks), thus avoiding the constraints associated with working with a large and homogeneous population. We further validate the PCP approach by studying a simple generative model, and by applying it to a data set of brain activity of healthy people.

The remainder of the text is organised as follows. Besides this introduction, Sections \ref{sec:CP} and \ref{sub:Probabilistic-Constraint-Program} respectively review the state of the art in constraint programming and its probabilistic version. Afterwards, the application of PCP is presented in Section \ref{sec:Results} for a data set of brain magneto-encephalographic recordings, and the advantages of PCP are discussed in Section \ref{sec:comparing}. Finally, some conclusions are drawn in Section \ref{sec:conclusions}.

\section{Constraint Programming}
\label{sec:CP}

A constraint satisfaction problem \cite{Mackworth1977} is a classical
artificial intelligence paradigm characterised by a set of variables
and a set of constraints, the latter specifying relations among subsets of these variables. Solutions are assignments of values to all variables that
satisfy all the constraints. 

Constraint programming is a form of declarative programming, in the
sense that instead of specifying a sequence of steps to be executed, it
relies on properties of the solutions to be found that are explicitly
defined by the constraints. A constraint programming framework must
provide a set of constraint reasoning algorithms that take advantage
of constraints to reduce the search space, avoiding regions inconsistent
with the constraints. These algorithms are supported by specialised
techniques that explore the specificity of the constraint model, such
as the domain of its variables and the structure of its constraints.

Continuous constraint programming \cite{Lhomme93IJ,benhamou94CLP}
has been widely used to model safe reasoning in applications where
uncertainty on the values of the variables is modelled by intervals
including all their possibilities. A Continuous Constraint Satisfaction
Problem (CCSP) is a triple $\left\langle X,D,C\right\rangle $, where
$X$ is a tuple of $n$ real variables $\left\langle x_{1},\cdots,x_{n}\right\rangle $,
$D$ is a Cartesian product of intervals $D(x_{1})\times\cdots\times D(x_{n})$
(a box), each $D(x_{i})$ being the domain of variable $x_{i}$,
and $C$ is a set of numerical constraints (equations or inequalities)
on subsets of the variables in $X$. A solution of the CCSP is a value
assignment to all variables satisfying all the constraints in $C$.
The feasible space $F$ is the set of all CCSP solutions within $D$.

Continuous constraint reasoning relies on branch-and-prune algorithms
\cite{Hentenryck97solvingpolynomial} to obtain sets of boxes that
cover exact solutions for the constraints (the feasible space $F$).
These algorithms begin with an initial crude cover of the feasible
space (the initial search space, $D$) which is recursively refined
by interleaving pruning and branching steps until a stopping criterion
is satisfied. The branching step splits a box from the covering into
sub-boxes (usually two). The pruning step either eliminates a box
from the covering or reduces it into a smaller (or equal) box maintaining
all the exact solutions. Pruning is achieved through an algorithm~\cite{granvilliers06algorithm} that combines constraint propagation
and consistency techniques~\cite{benhamou99revising}: each box is
reduced through the consecutive application of narrowing operators
associated with the constraints, until a fixed-point is attained. These
operators must be correct (do not eliminate solutions) and contracting
(the obtained box is contained in the original). To guarantee such
properties, interval analysis methods are used.

Interval analysis \cite{moore66interval} is an extension of real
analysis that allows computations with intervals of reals instead
of reals, where arithmetic operations and unary functions are extended
for interval operands. For instance, $[1,3]+[3,7]$ results in the
interval $[4,10]$, which encloses all the results from a point-wise
evaluation of the real arithmetic operator on all the values of the
operands. In practice these extensions simply consider the bounds
of the operands to compute the bounds of the result, since the involved
operations are monotonic. As such, the narrowing operator $Z\leftarrow Z\cap(X+Y)$
may be associated with constraint $x+y=z$ to prune the domain of
variable $z$ based on the domains of variables $x$ and $y$. Similarly,
in solving the equation with respect to $x$ and $y$, two additional
narrowing operators can be associated with the constraint, to safely
narrow the domains of these variables. With this technique, based
on interval arithmetic, the obtained narrowing operators are able
to reduce a box $X\times Y\times Z=[1,3]\times[3,7]\times[0,5]$ into
$[1,2]\times[3,4]\times[4,5]$, with the guarantee that no possible
solution is lost.

\subsection{Probabilistic Constraint Programming\label{sub:Probabilistic-Constraint-Program}}

In classical CCSPs, uncertainty is modelled by intervals that represent
the domains of the variables. Constraint reasoning reduces uncertainty,
providing a safe method for computing a set of boxes enclosing the
feasible space. Nevertheless this paradigm cannot distinguish between
different scenarios, and all combination of values within such enclosure
are considered equally plausible. In this work we use probabilistic
constraint programming \cite{ElsaPhD}, which extends the continuous
constraint framework with probabilistic reasoning, allowing to further
characterise uncertainty with probability distributions over the domains
of the variables.

In the continuous case, the usual method for specifying a probabilistic
model~\cite{halpern03reasoning} assumes, either explicitly or implicitly,
a joint probability density function (p.d.f.) over the considered
random variables, which assigns a probability measure to each point
of the sample space $\varOmega$. The probability of an event $\mathcal{H}$,
given a p.d.f. $f$, is its multidimensional integral on the region
defined by the event: 
\begin{equation}
P(\mathcal{H})=\int_{\mathcal{H}}f(\mathbf{x})d\mathbf{x}\label{eq:prob-event}
\end{equation}

The idea of probabilistic constraint programming is to associate a
probabilistic space to the classical CCSP by defining an appropriate
density function. A probabilistic constraint space is a pair $\left\langle \left\langle X,D,C\right\rangle ,f\right\rangle $,
where $\left\langle X,D,C\right\rangle $ is a CCSP and $f$ is a
p.d.f. defined in $\varOmega\supseteq D$ such that: $\int_{\mathcal{\varOmega}}f(\mathbf{x})d\mathbf{x=1}$.

A constraint (or a conjunction of constraints) can be viewed as an
event $\mathcal{H}$ whose probability can be computed by integrating
the density function $f$ over its feasible space as in equation (\ref{eq:prob-event}).
The probabilistic constraint framework relies on continuous constraint
reasoning to get a tight box cover of the region of integration $\mathcal{H}$,
and computes the overall integral by summing up the contributions of
each box in the cover. Generic quadrature methods may be used to evaluate
the integral at each box.

In this work, Monte Carlo methods \cite{hammersley64monte} are used
to estimate the value of the integrals at each box. The integral can
be estimated by randomly selecting $N$ points in the multidimensional
space and averaging the function values at these points. This method
displays $\frac{1}{\sqrt{N}}$ convergence, {\it i.e.} by quadrupling the
number of sampled points the error is halved, regardless of the number
of dimensions.

The advantages obtainable from this close collaboration between constraint pruning
and random sampling were previously illustrated in ocean colour remote
sensing studies \cite{carvalho13probabilistic}, where this approach
achieved quite accurate results even with small sampling rates. The
success of this technique relies on the reduction of the sampling
space, where a pure non-na\"ive Monte Carlo (adaptive) method is not
only hard to tune, but also impractical in small error settings.

\section{From brain activity to network models}
\label{sec:Results}

\begin{figure*}[!tb]
\begin{center}
{\center
\includegraphics[width=0.99\textwidth]{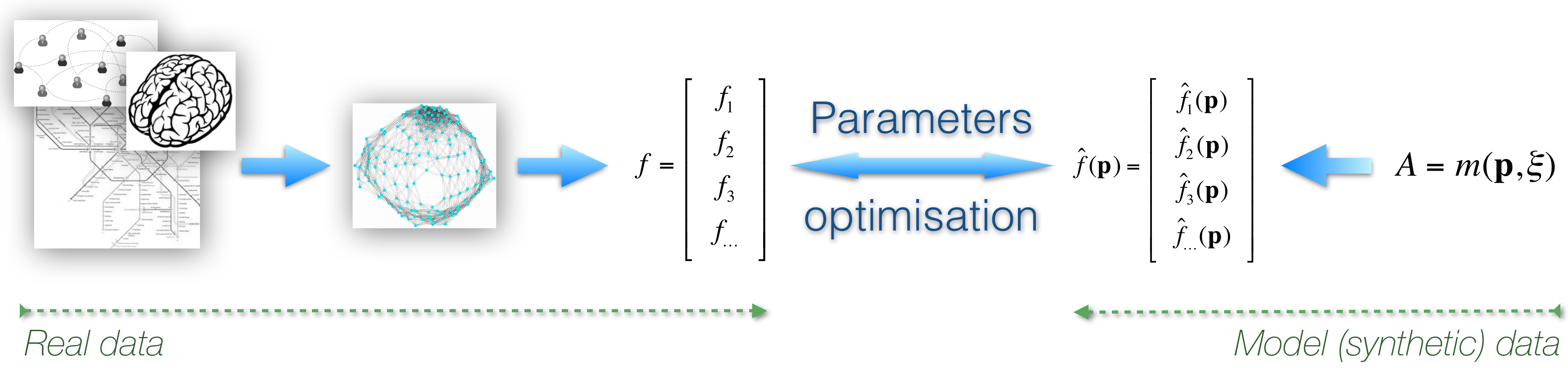}
}
\caption{Schematic representation of the use of generative models for analysing functional networks. $f$ and $\hat f$ respectively represent real and synthetic topological features, as the ones described in Sec. \ref{sec:reconstr}. Refer to Sec. \ref{sec:Results} for a description of all steps of the analysis.}\label{fig:01}
\end{center}
\end{figure*}

In order to validate the use of PCP for analysing the parameters space of a generative models, here we consider a set of magneto-encephalographic (MEG) recordings.
A series of preliminary steps are required, as shown in Fig. \ref{fig:01}. First, starting from the left, real brain data (or data representing any other real complex system) have to be recorded and encoded in networks, then transformed into a set of topological (structural) features. In parallel, as depicted in the right part, a generative model has to be defined: this allows to generate networks as a function of the model parameters, and extract their topological features. Finally, both features should be matched, {\it i.e.} the model parameters should be optimised to minimise the distance between the vectors of topological features of the synthetic and real networks.

\subsection{MEG data recording}
\label{sec:data}

Magneto-encephalographic (MEG) scans were obtained for $19$ right handed elderly and healthy participants, recruited from the Geriatric Unit of the Hospital Universitario San Carlos Madrid and the Centro de Prevenci\'on del Deterioro Cognitivo, Ayuntamiento de Madrid, Spain. Before the task execution, all participants or legal representatives gave informed consent to participate in the study. The study was approved by the local ethics committee.

Brain activity scans correspond to a modified version of the Sternberg's letter-probe task \cite{Maestu2001}, a standard task used to evaluate elders memory proficiency. MEG signals were recorded with a $254$ Hz sampling rate, using $148$-channel whole head magnetometer, confined in a magnetically shielded room (MSR). $35$ artefact-free epochs were randomly chosen from those corresponding to correct answers for each of participant.

\subsection{Networks reconstruction and evaluation}
\label{sec:reconstr}

Following the diagram of Fig. \ref{fig:01}, MEG recordings are converted in functional networks. Nodes, corresponding to MEG sensors and therefore to different brain regions, are pairwise connected when some kind of common dynamics is detected between the corresponding time series. Such relationship is assessed through Synchronization Likelihood (SL) \cite{Stam2002}, a metric able to detect {\it generalised synchronisation}, {\it i.e.} situations in which two time series react to a given input in different, yet consistent ways \cite{Yang1998}. It thus goes beyond simple linear correlations, as it is able to detect non-linear and potentially chaotic relations. Applying SL yields a correlation matrix $C \{ w _{ij} \}$ of size $148 \times 148$ (the number of sensors in the MEG machine) for each epoch available. In order to filter any kind of transient or noise specific to one epoch, the $35$ matrices corresponding to each subjects have been averaged: the final result is then a single weight matrix $\tilde{ C } \{ w _{ij} \}$ for each subject.

While a correlation matrix can readily be interpreted as a weighted fully-connected network, few metrics are available to describe the structure of such objects. It is then customary to apply a threshold, {\it i.e.} discard all links whose weight is not significant, and thus obtain an unweighted network. This presents several advantages. First of all, brain networks are expected to be naturally sparse, as increasing the connectivity implies a higher physiological cost. Furthermore, low synchronisation values may be the result of statistical fluctuations, {\it e.g.} of correlated noise; in such cases, deleting spurious links can only improve the understanding of the system. Lastly, a pruning can also help deleting indirect, second order correlations, which do not represent direct dynamical relationships. 

The final step involves the calculation of the topological metrics associated to each pruned network, {\it i.e.} the $f$s of Fig. \ref{fig:01}. Two have here been considered, representing two complementary aspects of brain information processing; their selection has been motivated by the generative model used afterwards (see Section \ref{sec:model}):

\begin{description}
	\item[Clustering coefficient.] The {\it clustering coefficient}, also known as {\it transitivity}, measures the presence of triangles in the network \cite{Newman2001}. Mathematically, it is defined as the relationship between the number of triangles and the number of connected triples in the network: $C = {{3N_\Delta  }} / {{N_3 }}$.
Here, a triangle is a set of three nodes with links between each pair of them, while a connected triple is a set of three nodes where each one can be reached from each other (directly or indirectly). From a biological point of view, the clustering coefficient represents how brain regions are locally connected, creating dense communities computing some information in a collaborative way.

\item[Efficiency.] It is defined as the inverse of the harmonic mean of the length of the shortest paths connecting pairs of nodes \cite{Latora2001}:

\begin{equation}
E = \frac{1}{{N\left( {N - 1} \right)}}\sum\limits_{i \ne j} { \frac{1}{d_{ij}} },
\end{equation}

$d_{ij}$ being the distance between nodes $i$ and $j$, {\it i.e.} the number of jumps required to travel between them. A low value of $E$ implies that all brain regions are connected by short paths.

\end{description}

\begin{figure*}[!tb]
\begin{center}
{\center
\includegraphics[width=0.4\textwidth]{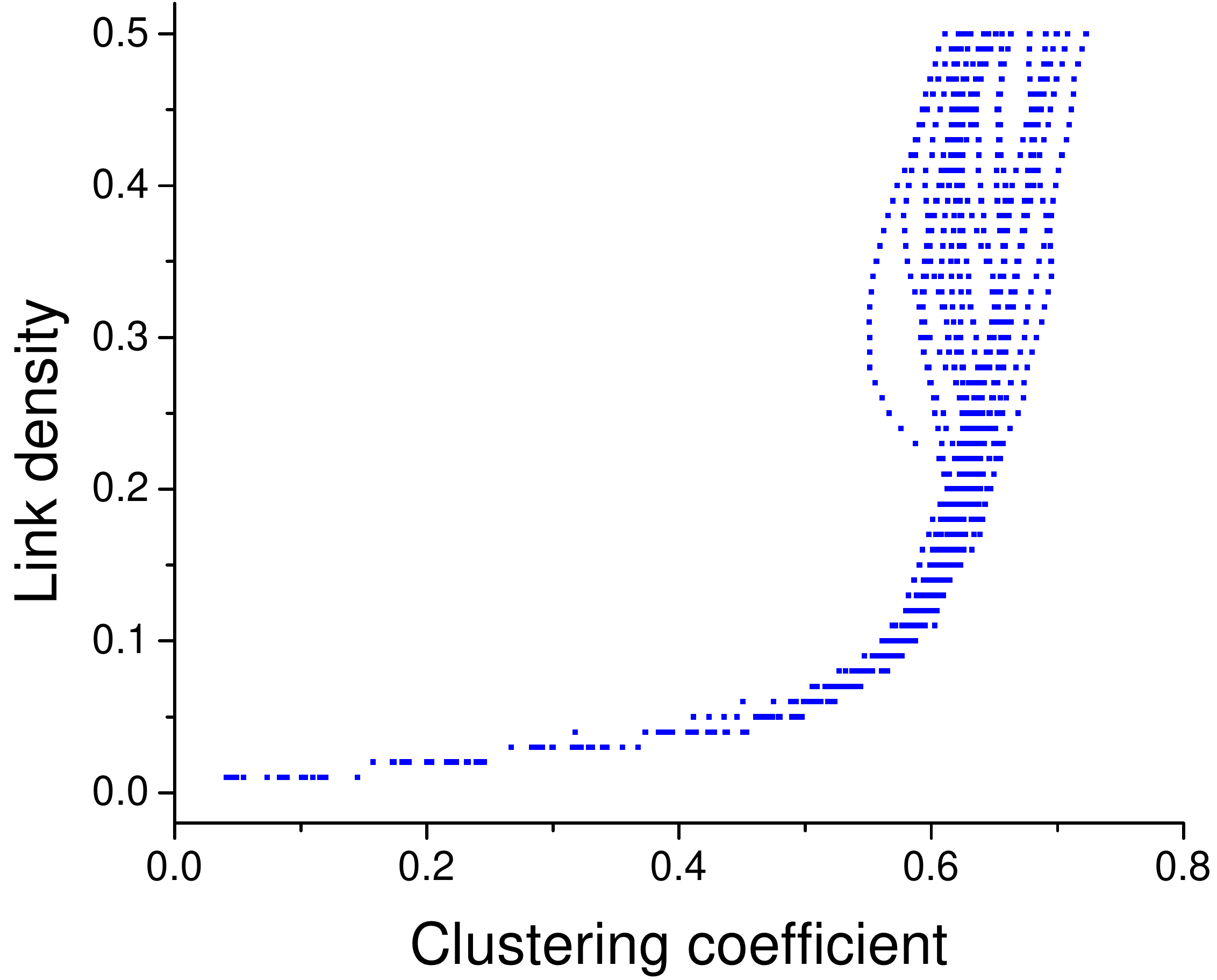}
\hspace{1cm}
\includegraphics[width=0.4\textwidth]{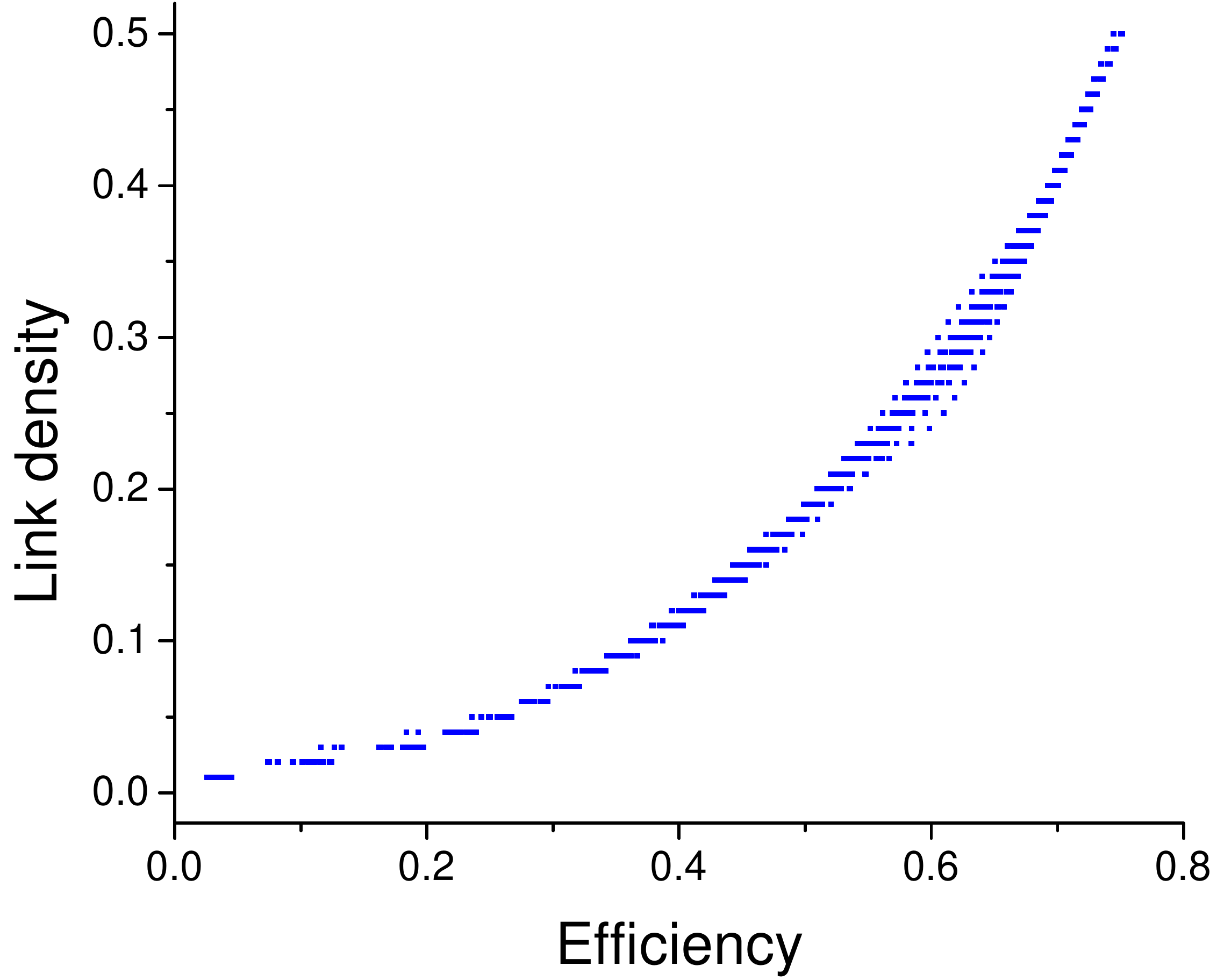}
}
\caption{Evolution of clustering coefficient (Left) and efficiency (Right) as a function of the link density for the $19$ functional brain networks reconstructed in Sec. \ref{sec:reconstr}. }\label{fig:02}
\end{center}
\end{figure*}

It has to be noticed how these two measures are complementary, the clustering coefficient and efficiency respectively representing the {\it segregation} and {\it integration} of information \cite{Tononi1994,Rad2012}. Additionally, both $C$ and $E$ are here defined as a function of the threshold $\tau$ applied to prune the networks - their evolution is represented in Fig. \ref{fig:02}.

\subsection{Generative model definition}
\label{sec:model}

Jumping to the right side of Fig. \ref{fig:01}, it is now necessary to define a generative model. As an example, we have here implemented a {\it Economical Clustering Model} model as defined in \cite{Vertes2012,Vertes2014}. Given two nodes $i$ and $j$, the probability of creating a connection between them is given by:

\begin{equation}
P_{i, j} \propto k _{i, j} ^ \gamma d _{i, j} ^ {- \eta}.
\label{eq:EconomicalClustering}
\end{equation}

$k _{i, j}$ is the number of neighbours common to $i$ and $j$, and $d _{i, j}$ is the physical distance between the two nodes. This model thus includes two different forces that compete to create links. On one side, $\gamma$  controls the appearance of triangles in the network, by positive biasing the connectivity between nodes having nearest neighbours in common; it thus defines the clustering coefficient and the appearance of computational communities. On the other side, $\eta$ accounts for the distance in the connection, such that long-range connections, which are biologically costly, are penalised.

\subsection{Parameters estimation through p-values}
\label{sec:par_std}

\begin{figure*}[!tb]
\begin{center}
{\center
\includegraphics[height=0.35\textwidth]{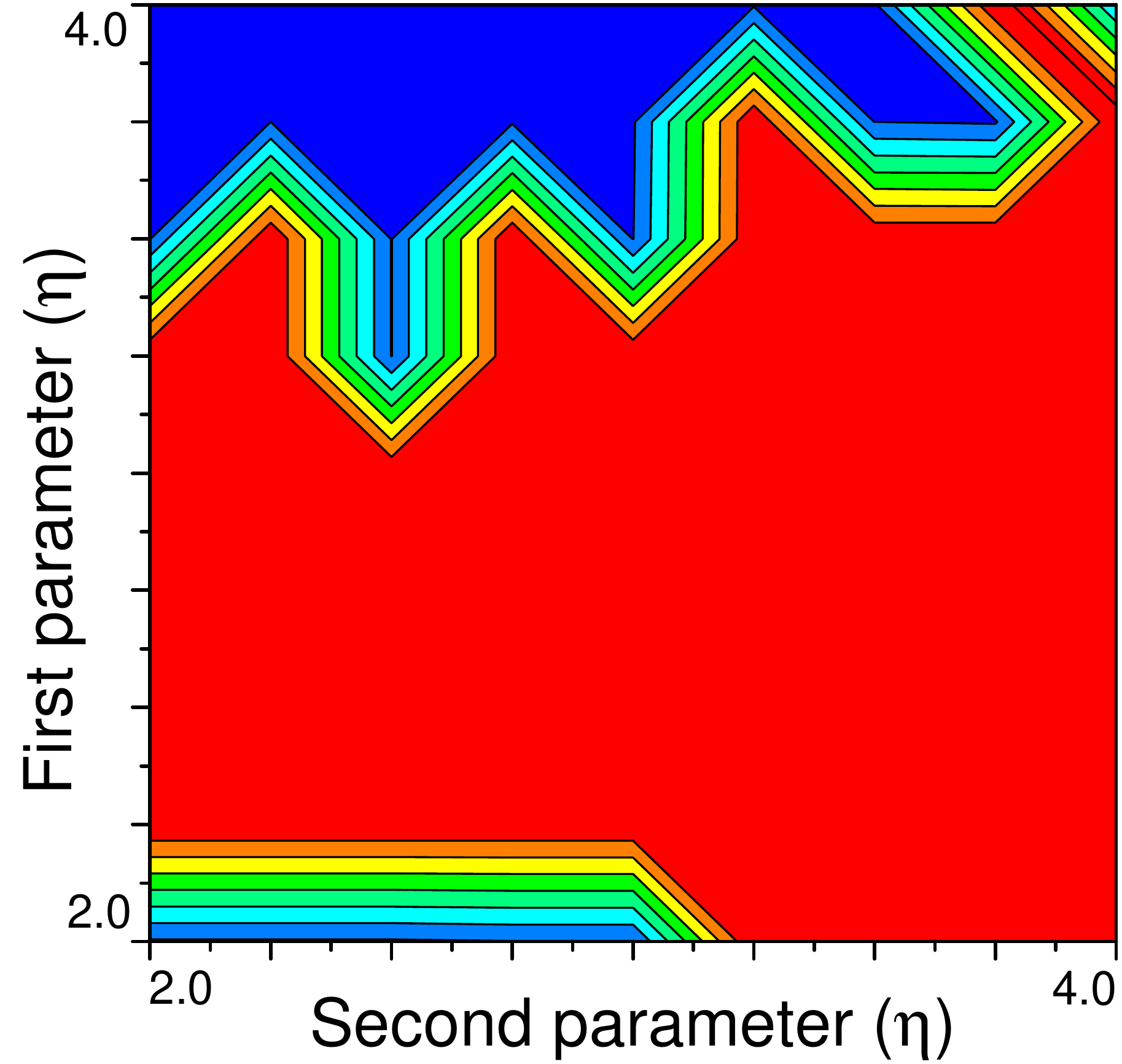}
\hspace{0.7cm}
\includegraphics[height=0.35\textwidth]{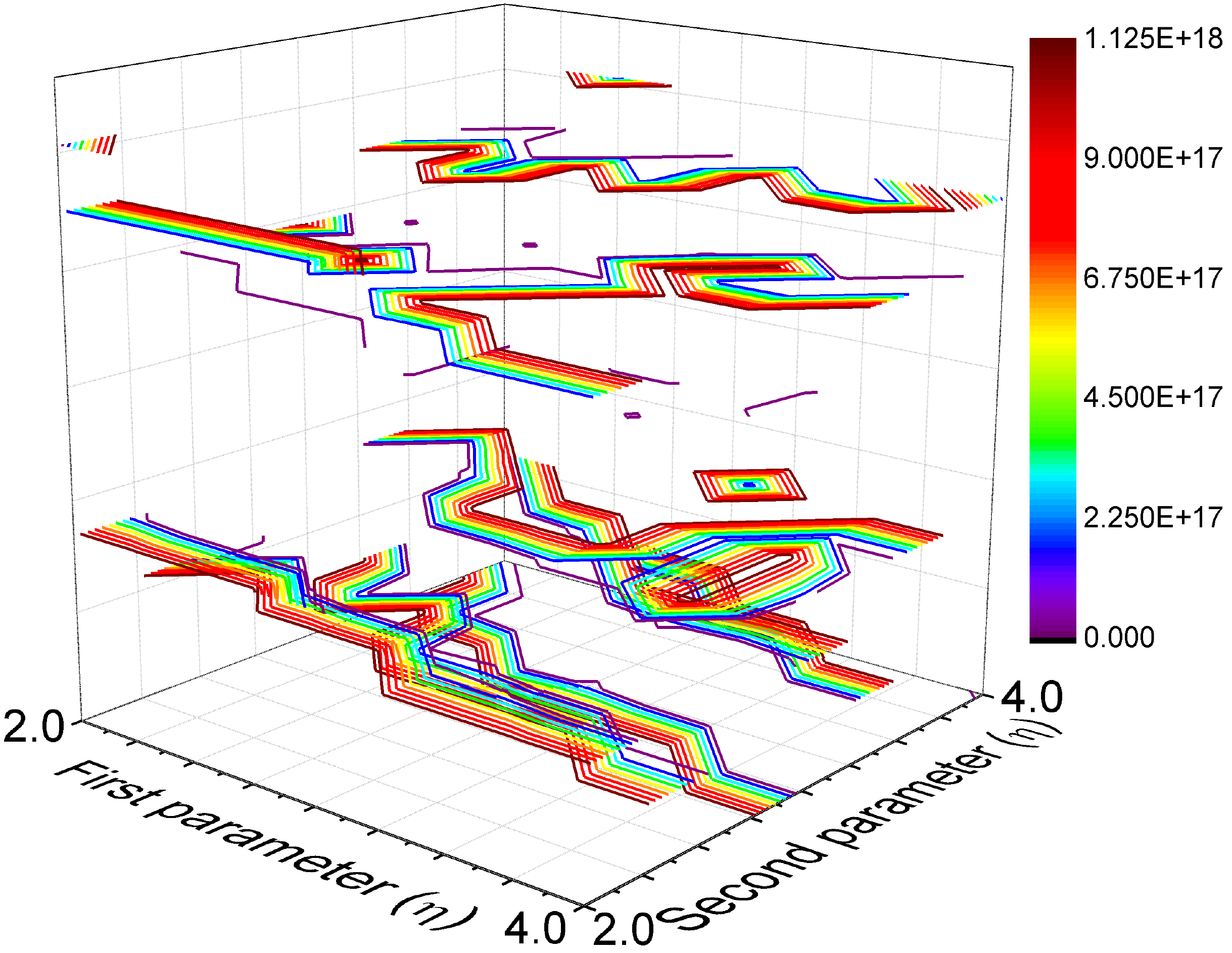}
}
\caption{(Left) Contour plot of the energy $E$ (see Eq. \ref{eq:energy1}) in the parameters space, for a link density of $0.3$. (Right) Energy contour plots for ten link densities, from $0.05$ (bottom) to $0.5$ (top); for the sake of clarity, only region outlines are visible.}\label{fig:03}
\end{center}
\end{figure*}

The problem is now identifying the best values of $\gamma$ and $\eta$ that permit recovering the topological properties obtained in Sec. \ref{sec:reconstr} for the experimental brain networks. As an example of a standard $p$-value based mechanism, we here use a simplified version of the energy function proposed in Refs. \cite{Vertes2012,Vertes2014}:

\begin{equation}
E = {1} / { \prod \limits_i P_i}.
\label{eq:energy1}
\end{equation}

$P_i$ represents the $p$-value of the Kolmogorov-Smirnoff (K-S) test between the distributions estimated from the model and experimental networks, and $i$ runs over all topological metrics. As just two topological properties are here studied, the previous formula simplifies to: $E = {1} / { ( P_{E} \cdot P_{C} ) }.$

For each considered value of $\gamma$ and $\eta$, a set of networks have been generated according to the model of Eq. \ref{eq:EconomicalClustering}; their topological features extracted; and the resulting probability distribution compared with the distribution corresponding to the real networks, through a K-S test.

Fig. \ref{fig:03} presents the result of plotting the energy evolution in the parameters space. Specifically, Fig. \ref{fig:03} Left reports the evolution of the energy for a link density of $0.3$. It can be noticed that a large portion of the space, constructed around the values of $\gamma$ and $\eta$ suggested in \cite{Vertes2012}, maximises the energy. Fig. \ref{fig:03} Right represents the same information for ten different link densities, from $0.05$ (bottom part) to $0.5$ (upper part).

\subsection{Parameters estimation through Probabilistic Constraint Programming}
\label{sec:par_pcp}

As an alternative solution, the previously described PCP method is here used to recover the shape of the parameters space. 
Two preliminary steps have to be completed: first, reconstruct a set of synthetic networks using the generative model of Eq. \ref{eq:EconomicalClustering}, for different $\gamma$ and $\eta$ values, and extract their topological characteristics; and second, obtain approximated functions describing the evolution of the topological metrics as a function of the model parameters, {\it i.e.} $C = \tilde{f_{C}}(\gamma, \eta)$ and $E = \tilde{f_{E}}(\gamma, \eta)$.
Afterwards, each
observed feature $o_{i}$ is modelled as a function $f_{i}$ of the
model parameters plus an associated error term $\epsilon_{i}\sim\mathcal{N}\left(\mu=0,\sigma^{2}\right)$:
\[
	o_{i}=f_{i}\left(\gamma,\eta\right)+\epsilon_{i}
\]

For $n$ observations, a probabilistic constraint space is considered
with random variables $\gamma$ and $\eta$, a set of constraints
$C$,
\[
C=\left\{ -3\sigma\leq o_{i}-f_{i}\left(\gamma,\eta\right)\leq3\sigma|1\leq i\leq n\right\} 
\]
$3\sigma$ being chosen to keep the error within reasonable bounds, and the joint p.d.f. $f$,
\begin{equation}
f\left(\gamma,\eta\right)=\prod_{i=1}^{n}g\left(o_{i}-f_{i}\left(\gamma,\eta\right)\right)\label{eq:error-pdf}
\end{equation}
where $g$ is the normal distribution with $0$ mean and standard
deviation $\sigma$.

To compute the probability distribution of the random variables $\gamma$
and $\eta$, a grid is constructed over their domains and a branch-and-prune
algorithm is initially used to obtain a grid box cover of the feasible
space (where each box belongs to a single grid cell). Then, for each
box in the cover, a Monte Carlo method is used to compute its contribution
to equation (\ref{eq:prob-event}) with the p.d.f. defined in equation
(\ref{eq:error-pdf}). The probability of the respective cell is updated
accordingly and normalised in end of the process.

\begin{figure*}[!tb]
\begin{center}
{\center
\includegraphics[width=0.65\textwidth]{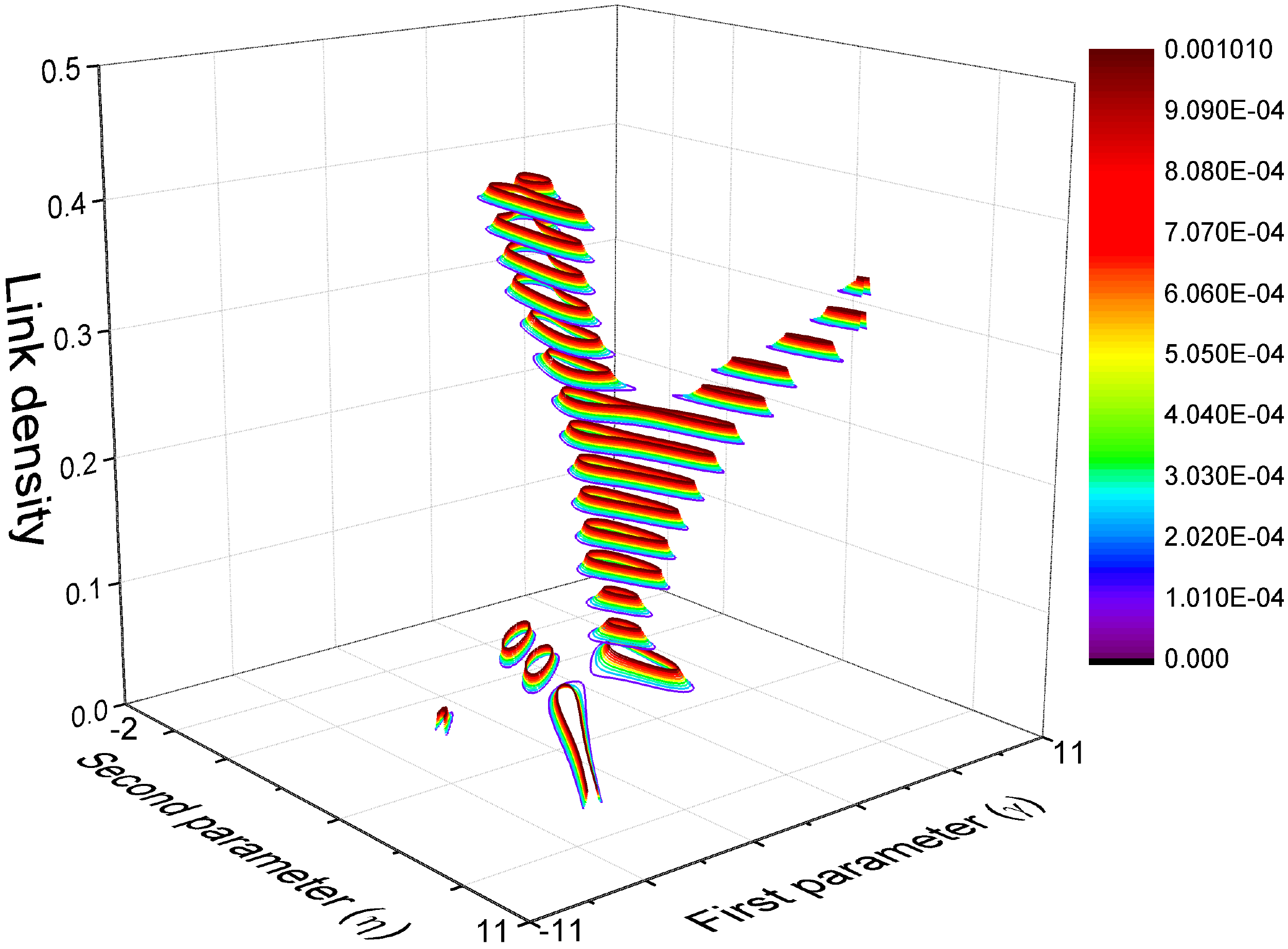}
}
\caption{Contour plot of the parameters space, as obtained by the PCP method, for the whole population of subjects and as a function of the link density. The colour of each point represents the normalised probability of generating topologically equivalent networks.}\label{fig:04}
\end{center}
\end{figure*}

Fig. \ref{fig:04} reports the results obtained, {\it i.e.} the probability of obtaining networks with the generative model which are compatible with the real ones, as a function of the two parameters $\gamma$ and $\eta$, and as a function of the link density. In the next Section, both approaches and their results are compared.

\section{Comparing p-value and Probabilistic Constraint Programming}
\label{sec:comparing}

Results presented in Sec. \ref{sec:par_std} and \ref{sec:par_pcp} allow comparing the $p$-value and PCP methods, and highlight the advantages that the latter presents over the former.

\begin{figure*}[!tb]
\begin{center}
{\center
\includegraphics[width=0.99\textwidth]{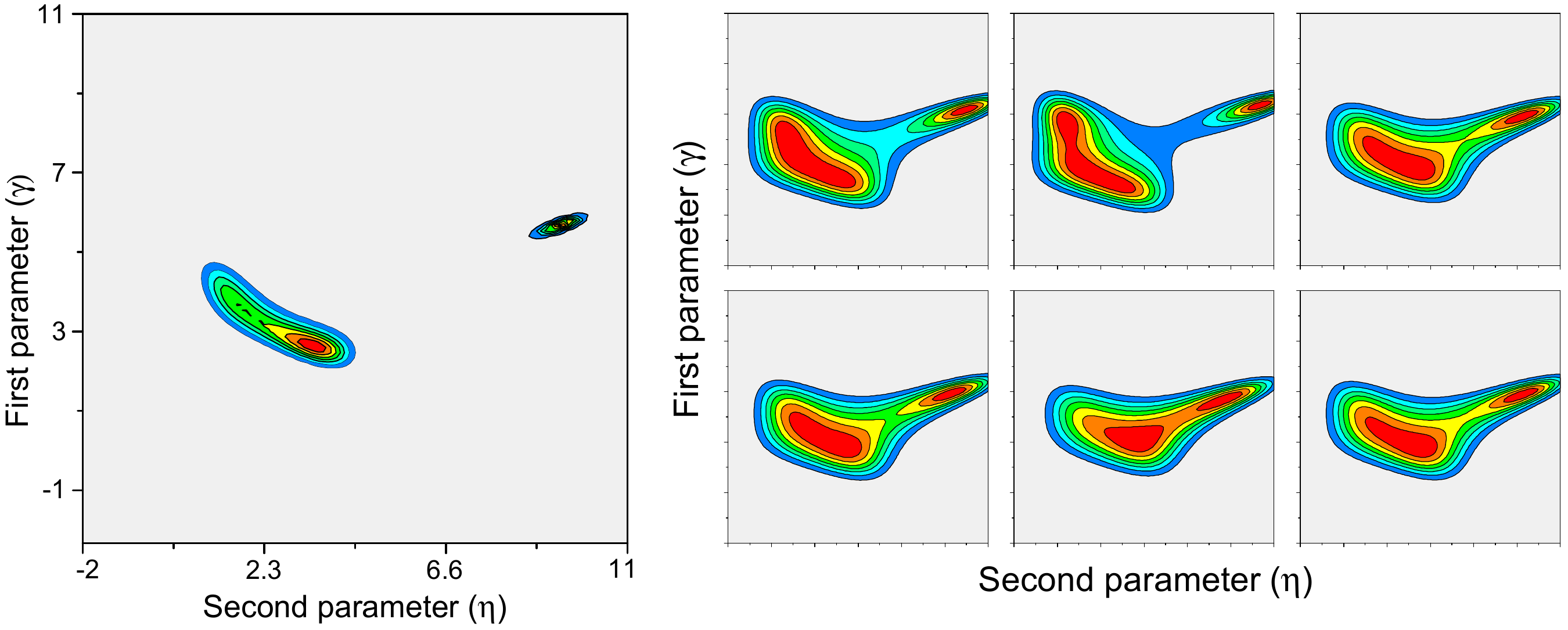}
}
\caption{(Left) Parameters space, as obtained with the PCP method, for a link density of $0.3$ and for the whole studied population. (Right) Parameters space for six subjects. The scale of the right graphs is the same as the left one; the colour scale is the same of the one of Fig. \ref{fig:04}.}\label{fig:05}
\end{center}
\end{figure*}

The extremely high computational cost of analysing the parameters space by means of K-S tests seldom allows a full characterisation of such space. This is due to the fact that, for any set of parameters, a large number of networks have to be created and characterised. Increasing the resolution of the analysis, or enlarging the region of the space considered, increases the computational cost in a linear way. This problem is far from being trivial, as, for instance, the networks required to create Fig. \ref{fig:03} represents approximatively $3$ GB of information and several days of computation in a standard computer.
Such computational cost implies that it is easy to miss some important information. Let us consider, for instance, the result presented in Fig. \ref{fig:03} Left. The shape of the iso-lines suggests that the maximum is included in the region under analysis, and that no further explorations are required - while Figs. \ref{fig:04} and \ref{fig:05} prove otherwise.

On the other hand, estimating the functions $\tilde{f_{C}}$ and $\tilde{f_{E}}$ requires the creation and analysis of a constant number of networks, independently on the size of the parameters space. The total computational cost drops below the hour in a standard computer, implying a 3 orders of magnitude reduction.
This has important consequences on the kind of information one can obtain. Fig. \ref{fig:05} Left presents the same information as Fig. \ref{fig:03} Left, but calculated by means of PCP over a larger region. It is then clear that the maximum identified in Fig. \ref{fig:03} is just one of the two maxima presents in the system.

The second important advantage is that, while the PCP can yield results for just one network or subject, a $p$-value analysis requires a probability distribution. It is therefore not possible to characterise the parameters space for just one subject, but only for a large population.
Fig. \ref{fig:05} Right explores this issue, by showing the probability evolution in the parameters space for six different subjects. It is interesting to notice how subjects are characterised by different shapes in the space. This allows a better description of subjects, aimed for instance at detecting differences among them.

\section{Conclusions}
\label{sec:conclusions}

In this contribution, we have presented the use of Probabilistic Constraint Programming for optimising the parameters of a generative model, aimed at describing the mechanisms responsible for the appearance of some given topological structures in real complex networks. As a validation case, we have here presented the results corresponding to functional networks of brain activity, as obtained through MEG recordings of healthy people.

The advantages of this method against other customary solutions, {\it e.g.} the use of $p$-values obtained from Kolmogorov-Smirnoff tests, have been discussed. First, the lower computational cost, and especially its independence on the size of the parameters space and on the resolution of the analysis. This allows a better characterisation of such space, reducing the risk of missing relevant results when multiple local minima are present. Second, the possibility of characterising the parameters space for single subjects, thus avoiding the need of having data for a full population. This will in turn open new doors for understanding the differences between individuals: as, for instance, for the identification of characteristics associated to specific diseases in diagnosis and prognosis tasks.

\end{document}